\documentclass{mnras}
\usepackage{xcolor, graphicx}

\begin{document}

\title[GC system gradients]{Metallicity gradients in the globular cluster systems of early-type galaxies: In-situ and accreted components?}
\author[Forbes \& Remus]{Duncan A. Forbes$^{1}$\thanks{E-mail:
dforbes@swin.edu.au},  Rhea-Silvia Remus$^{2,3}$
\\
$^{1}$Centre for Astrophysics \& Supercomputing, Swinburne
University, Hawthorn VIC 3122, Australia\\
$^{2}$Universit\"ats-Sternwarte M\"unchen, Scheinerstr. 1, D-81679 M\"unchen, Germany\\
$^3$Canadian Institute for Theoretical Astrophysics, 60 St. George Street, University of Toronto, Toronto ON M5S 3H8, Canada
}


\pagerange{\pageref{firstpage}--\pageref{lastpage}} \pubyear{2002}

\maketitle

\label{firstpage}

\begin{abstract}

Massive early-type galaxies typically have two subpopulations of globular clusters (GCs) which often reveal radial colour (metallicity) gradients. Collating gradients from the literature, we show that the gradients in the metal-rich and metal-poor GC subpopulations are the same, within measurement uncertainties, in a given galaxy. Furthermore, these GC gradients are similar in strength to the {\it stellar} metallicity gradient of the host galaxy. At the very largest radii (e.g. greater than 8 galaxy effective radii) there is some evidence that the GC gradients become flat with near constant mean metallicity. 
Using stellar metallicity gradients as a proxy, we probe the assembly histories of massive early-type galaxies with hydrodynamical simulations from the Magneticum suite of models. In particular, we measure the stellar metallicity gradient for the in-situ and accreted components over a similar radial range as those observed for GC subpopulations. We find that the in-situ and accreted stellar metallicity gradients are similar but have a larger scatter than the metal-rich and metal-poor GC subpopulations gradients in a given galaxy.  We conclude that although metal-rich GCs are predominately formed during the in-situ phase and metal-poor GCs during the accretion phase of massive galaxy formation, they do not have a strict one-to-one connection. 

\end{abstract}


\section{Introduction}

A fundamental feature of most well-studied globular cluster (GC) systems is that they are bimodal in colour and metallicity (Brodie \& Strader 2006; 
Usher et al. 2012; Brodie et al. 2012)  consisting of 
blue (metal-poor) and red (metal-rich) GC subpopulations. In early-type galaxies 
the red GCs have long been associated with the underlying starlight as they share many properties in common. For example:\\ 
$\bullet$ the galaxy stellar surface brightness profile beyond the inner core region is well-matched by the surface density profile of red GCs (Bassino et al. 2006; Strader et al. 2011; Forbes et al. 2012)\\
$\bullet$ the shape of the red GC's azimuthal distribution follows that of the galaxy ellipticity  (Park \& Lee 2013; Kartha et al. 2014). \\
$\bullet$  the kinematics of the galaxy stars typically matches those of the red GCs more closely than the blue ones (Pota et al. 2013). \\
$\bullet$ the metallicity of the stars and red GCs are well-matched, at least in galaxy inner regions (Geisler et al. 1996; Lee et al. 1998;  Forbes \& Forte 2001; Lee et al. 2008). \\

These trends suggest a close connection in the  formation/assembly process for the red GCs and the underlying starlight. Indeed, several early works associate red GCs with {\it in-situ} formation contemporaneously with the host (spheroidal) galaxy (Forbes et al. 1997; Cote et al. 1998; Harris et al. 2000; Forbes \& Forte 2001). 

On the other hand, the blue GCs  are more radially extended, do not share the same kinematics, have different azimuthal distributions and typically have 1/10th, or less, the metallicity of the underlying starlight of their host galaxy.  They have been associated with an {\it accreted} component (e.g. Cote et al. 1998). 

In contemporary models, within the context of hierarchical galaxy assembly, the metallicity of GCs are set both by the mass of the host galaxy and the redshift of formation (Tonini 2013; Li \& Gnedin 2014; Choksi et al. 2018). So at the earliest epochs, GCs were largely metal-poor, while at later times, metal-rich GCs formed 
in-situ within relatively large host galaxies. Thus, as most accretion events are small satellites they contribute mostly metal-poor GCs.
This picture is supported by arguments that the majority of the 
outer halo GCs in the Milky Way have been accreted from dwarf satellites (e.g. Mackey \& Gilmore 2003;  Boylan-Kolchin 2017) with the Sgr dwarf being the most dramatic example of GC accretion (Law \& Majewski  2010; Forbes \& Bridges 2010). 

Two near-linear scaling relations indicate a strong connection between GC systems and the dark matter content of their host galaxy; these are  the total mass of a GC system with its host galaxy halo mass (Spitler \& Forbes 2009; Hudson et al. 2014) and the size of a GC system with the virial radius of the galaxy dark matter halo (Forbes 2017; Hudson \& Robison 2018). There is evidence that the blue GC are more closely linked to the dark matter content than the red GCs (Harris et al. 2015; Forbes et al. 2016). 
Together this suggests a relation between GC systems and galaxy dark matter halos that is established at early epochs, with galaxies growing by accumulating GCs and dark matter in tandem. 

Massive early-type galaxies (ETG) are thought to form in two-phases, with more massive galaxies
acquiring a larger fraction of accreted material from satellites (Oser
et al. 2010; Naab et al. 2014) than low mass galaxies. This accreted material serves to
build-up the halo of the host galaxy, so that the more massive galaxies 
tend to have shallower surface brightness
profiles (Pillepich et al. 2014) and shallower metallicity profiles (Hirschmann et al. 2015; Cook et al. 2016). In-situ formed stars and stars accreted in major mergers are more centrally located than stars accreted from minor mergers (Rodriguez-Gomez et al. 2016; Karademir et al. 2018). 

Observationally it is difficult to measure {\it stellar} metallicity gradients to large radii (a notable example is the 13.5hr integration on NGC 4889 to reach 4 effective radii by Cocatto et al. 2010). However such radii are occupied by globular clusters.  The radial colour (metallicity) gradients of the individual GC subpopulations have been traced to large galactocentric radii in a number of ETGs. We note that examining the {\it total} GC system colour gradient simply reflects the changing proportion of red and blue GCs with radius.
Given the picture of in-situ formed red GCs and ex-situ formed blue GCs summarised above, one might expect the radial metallicity gradients of the two GC subpopulations to be linked to the two phases of galaxy formation. 

Here we combine studies of GC systems in massive ETGs from the 
SLUGGS survey (Brodie et al. 2014) 
with those from the literature for which colour gradients have been measured for the blue and red 
GC subpopulations separately to large galactocentric radii. We convert these colour gradients to metallicity gradients and examine their properties. We find evidence for the same gradients in both the blue and red GC subpopulations in a given galaxy and 
discuss the implications of our findings in the context of two-phase galaxy formation using recent galaxy hydrodynamical simulations.

\begin{table*}
\centering
{\small \caption{Galaxy and globular cluster system properties}}
\begin{tabular}{@{}ccccccccc}
\hline
\hline
Galaxy & log M$_{\ast}$ & Blue grad & Red grad & Range & Ref. & n  & SLUGGS & Telescope\\
$\rm [NGC]$ & [M$_{\odot}$] & [dex/dex] & [dex/dex] & [R$_e$] & & & &\\
(1) & (2) & (3) & (4) & (5)  & (6) & (7) & (8) & (9)\\
\hline
1399 & 11.4 	  &    -0.12 (0.05) & -0.10 (0.05) & 0.2-5.5 & L11	&	4.5 & -- & HST \\
1407 & 11.6  	   &	 -0.22 (0.03) & -0.24 (0.07) & 0.5-8.5 & Fo11	&	4.9 & $\surd$  & Subaru\\
1407	 & 11.6	&	-0.075 (0.079)	&     -0.061 (0.064) & 0.5-5 & H09a & 4.9  & -- & HST\\

3115 & 10.9 &		-0.17 (0.03) & -0.24 (0.06) & 1-10 & A11	&	4.7 & $\surd$ & Subaru\\
3115 &	10.9	&      -0.27 (0.06) & -0.10 (0.10) & 1-8 & F11		&	4.7 & -- &  Gemini\\

3258      & 11.0	&	-0.075 (0.057)	 &    -0.068 (0.054) & 0.5-5 & H09a &	5.0 & -- &  HST\\
3268	 & 11.1	&	-0.095 (0.004)	 &    -0.005 (0.057) & 0.5-5 & H09a &	5.1 & -- & HST\\
3348	 & 11.1	&	-0.039 (0.079)	&     -0.082 (0.086) & 0.5-5 &H09a	 & 5.3 & -- & HST\\

3607 & 11.4 	&	-0.24 (0.05 & -0.12 (0.05) & 1.3-7.5 & K16	&	5.3 & $\surd$ & Subaru\\
3923  & 11.3 &	      -0.18 (0.07 & -0.17 (0.08) & 0.5-5.5 & F11	&		3.8 & -- & Gemini\\

4278 & 11.0	&	-0.29 (0.03) & -0.24 (0.06) & 0.5-8.5 & U13	& 6.2 & $\surd$ & Subaru+HST\\
4365 & 11.5 	&	-0.19 (0.01) & -0.22 (0.03) & 1-11 & B12	& 4.9 & $\surd$ & Subaru+HST\\
4472 & 11.6 	   &   -0.08 (0.04) & -0.09 (0.05) & 0.2-3 &L11		&	6.0 & -- & HST\\
4472  & 11.6        &   -0.15 (0.03) & -0.12 (0.06) & 0.5-5.5 & G96	&	6.0 & -- & KPNO\\
4486 & 11.6 	 &     -0.13 (0.02) & -0.13 (0.03) & 0.2-14 & L11	&	5.1 & -- & HST\\
4486 & 11.6 	&	-0.09 (0.02) & -0.12 (0.02) & 1-10 & H09b	& 5.1 & -- & CFHT\\
4486 & 11.6 	&	-0.17 (0.07) & -0.17 (0.05) & 0.2-3.5 & F12	& 5.1 & -- & HST\\
4594 & 11.4	&	-0.18 (0.04) & -0.16 (0.04) & 0.6-6 & HR14	& 3.2 & -- & KPNO\\
4649 &	11.6 	   &   -0.003 (0.038) &	-0.05 (0.02) & 0.5-5.5 & F11 &		4.6 & -- & Gemini\\
4696      & 11.6 &		-0.086 (0.032)	 &    -0.025 (0.041) & 0.5-5 & H09a & 7.1 & -- & HST\\
7626      & 11.4 	&	-0.136 (0.052) &	     -0.154 (0.059) & 0.5-5 & H09a	& 5.6 & -- & HST\\
BCGs       & 11.4 &      	-0.086 (0.015)	&     -0.084 (0.018) & 0.5-5 & H09a & 5.5 & -- & HST\\
ETGs & $>$11 & -0.08 (0.09) & -0.07 (0.08) & 0.6-14 & P15 & -- & $\surd$ & Keck\\

\hline
\end{tabular}
\begin{flushleft}
{\small 

Notes: columns are (1) galaxy name (BCGs = average of 6 BCGs, ETGs = average of 6 massive ETGs),   (2) stellar mass from F17 or corrected 2MASS 
K magnitude and M/L$_K$ = 1 (typical uncertainty is $\pm$0.1 dex),  (3) and (4) blue and red GC metallicity gradients and uncertainty, (5) radial range in units of effective radii, (6) original colour gradient reference, H09a  = Harris (2009a), F11 = Faifer et al. (2011), L11 = Liu et al. (2011), G96  = Geisler et al. (1996), Fo = Forbes et al. (2011), A11  = Arnold et al. (2011), K16 = Kartha et al. (2016), U13 = Usher et al. (2013), B12 = Blom et al. (2012), H09b = Harris (2009b), F12 = Forte et al. (2012), HR14 = 
Hargis \& Rhode (2014), P15 = Pastorello et al. (2015), (7) Sersic slope n value for galaxy surface brightness profile, (8) SLUGGS galaxy, (9) telescope used.

}

\end{flushleft}
\end{table*}

\section{The sample}

We select early-type galaxies (ETGs) for which {\it both} blue and red GC subpopulation radial gradients have been measured  in a given galaxy to large radii (i.e. $\ge$3 R$_e$). 
Our sample consists of 5 ETGs imaged by the SLUGGS survey and 16 others from the literature.
We also include the average value derived for six BCGs by Harris (2009a), and the average value for 
six high mass ETGs of the SLUGGS survey by Pastorello et al. (2015). 

Measured GC colour gradients were converted 
into metallicity gradients using a variety of different
transformations. Here we take their measured colour gradients and convert them
into a [Z/H] metallicity gradient using the linear transformations
from Usher (2014). This gives slightly different gradients
to those reported by the different studies but puts them on a more
homogeneous basis.  The exceptions are the studies of Geisler et
al. (1996) who measured C-T colour gradients and reported metallicity
gradients, Hargis \& Rhode (2014) who worked with B--R colours and
Forte et al. (2012) who only reported metallicity gradients. The
colour-metallicity transformation we use for the remaining studies are:\\

\noindent
$[Z/H]$ = 3.49 $\times$ (g--i) - 4.03\\
$[Z/H]$ = 2.56 $\times$ (g--z) - 3.50\\
$[Z/H]$ = 2.27 $\times$ (B--I) - 4.68\\

The metallicity gradients listed in Table 1 include several values for some galaxies, measured by different studies 
over different radii. They are NGC 1407 (2 measurements), 
NGC 3115 (2 measurements), NGC 4472 (2 measurements) and NGC 4486 (3 measurements). In general they agree within 
the reported uncertainities. All gradients are measured over different radii, although typically 
they extend to $\ge$ 5 R$_e$.\\

We also measure the gradient in the stellar surface brightness
profile (in terms of intensity/per unit area) for the galaxies listed in Table 1 over the same radial
interval as the measured GC gradients. The gradient is a power-law fit to the 
Sersic parameters  of each galaxy's surface brightness profile i.e. 
effective radius (R$_e$) and Sersic slope (n), which is a good approximation over a limited radial range. 
Most galaxies in our sample have R$_e$ and n from Sersic fits 
available in Forbes et al. (2017). We supplement this using the relationship between 
absolute luminosity and Sersic parameters from Graham (2013). 

\section{Results from Observations}

\begin{figure*}
      \includegraphics[angle=-90, width=0.95\textwidth]{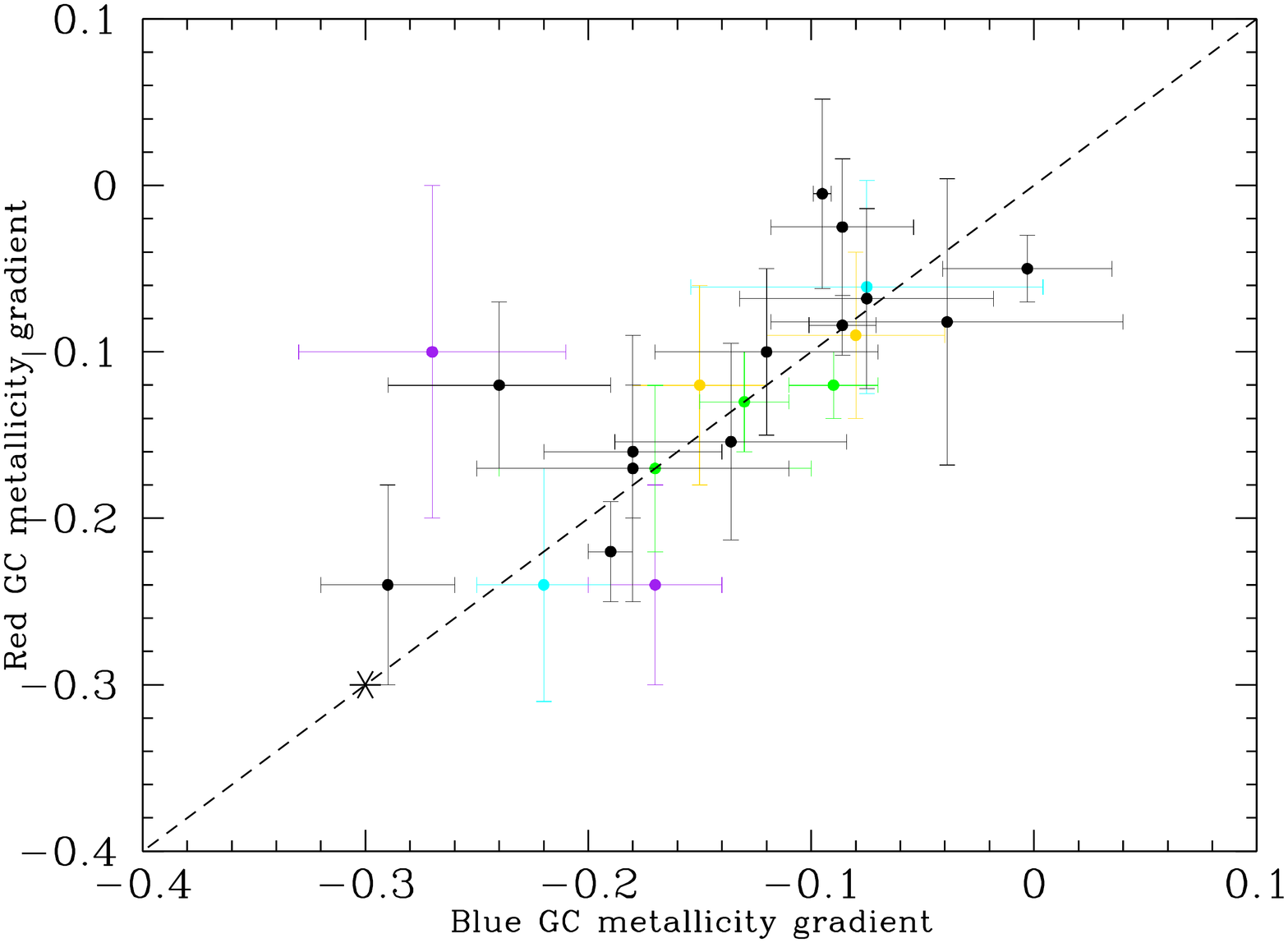}
	    \caption{\label{fig:unity} Globular cluster metallicity gradients, with units of dex per dex, for blue and red GC subpopulations. 
The solid line shows a 1:1 relation. Gradients from different studies for the same galaxy are shown with the same coloured symbol (i.e green for NGC 4486, purple for NGC 3115, cyan for NGC 1407 and yellow for NGC 4472), whereas black symbols represent galaxies for which there is only one measurement of the gradient. 
The Milky Way GC system is shown by an asterisk. For a given galaxy, the blue and red GC metallicity gradients are the same within the uncertainties.  
}

\end{figure*} 

\subsection{Metal-poor vs metal-rich  GC metallicity gradients}

GC systems typically extend to $\sim$15 times the effective radius (R$_e$) of their host galaxy (Kartha et al. 2016)  and have been confirmed in some cases out to 20 R$_e$ (Alabi et al. 2016). 
The GC gradients discussed above are typically measured out to 5-8 R$_e$ (which we arbitrarily call the `inner halo').
These `inner halo' colour gradients are transformed into metallicity and tabulated in Table 1. In Fig. ~\ref{fig:unity} we show  the metallicity gradient for the metal-rich GCs against that for the metal-poor GCs for our sample of massive ETGs. We note that some galaxies are shown several times (e.g. NGC 4486 has 3  literature measurements available) as highlighted in the plot. 
In addition to the ETGs from Table 1, we include the gradients quoted for the 
metal-poor and metal-rich GC subpopulations of the Milky Way within 10 kpc (Harris 2001).
Within the uncertainties, the data are consistent with a one-to-one relation between the metal-rich and metal-poor GC metallicity gradients. This supports the claim of 
Harris (2009a), based on half a dozen galaxies, that both GC subpopulations reveal the same metallicity gradient within a given galaxy.
We have examined the distribution of GC metallicity gradients with host galaxy stellar mass but find no obvious trend. 

It is worth noting, that although the radial gradients discussed above imply that the mean metallicity of the metal-rich subpopulation declines with galactocentirc radius it does not decline enough to reach the levels of the metal-poor subpopulation. In other words, the radial metallicity gradient is not steep enough to explain the mean metallicity of the metal-poor GCs (see Forbes et al. 2011 for an example). So unless the radial gradients were significantly steeper in the past, the metallicity of the metal-poor subpopulation is set by some other process, e.g. the mass of the host galaxy and the epoch of formation. This further suggests that metal-poor GC formation occurs at early times in low mass galaxies (which are subsequently accreted by larger galaxies).

\subsection{GC vs stellar metallicity gradients}

In an early photometric study of a GC system and its host galaxy (NGC 4472) using the same metallicity-sensitive filters, Geisler et al. (1996) found that the red GCs had a similar {\it mean} metallicity to that of the underlying starlight over radii for which reliable halo colours could be measured. 
Forbes \& Forte (2001) looked at the mean colours of several ETGs and their red GCs, matching their coverage in galactocentric radii, and found a near one-to-one relation. Several studies since then (e.g. Lee et al. 2008 on NGC 4649) have also found that the  red GCs in ETGs have a similar mean colour to the central stars of their host galaxy. More recently, in a study of half a dozen massive ETGs, Goudfrooij \& Kruijssen (2013)  found a small, but systematic, offset in the mean colour of the host galaxy and its red GCs (they interpreted this difference as due to a bottom-heavy IMF in the galaxy stars). 


As well as similar mean colours/metallicities, GC systems and the underlying galaxy starlight may have similar metallicity gradients.  For example, Geisler et al. (1996) found that the red GC metallicity gradient (--0.12 $\pm$ 0.06 dex per dex), blue GC gradient (--0.15 $\pm$ 0.03 dex per dex) and the stellar metallicity gradient within $\sim$4 R$_e$ (--0.14 dex per dex) for NGC 4472 were the same within the uncertainties. Since then, only a few studies of GC and stellar metallicity gradients have been published using the same method over a large common radial range (the difficulty being that the galaxy colour becomes increasingly hard to measure at the large radii occupied by the bulk of the GCs). Two massive ETGs warrant mention here -- NGC 1407 and NGC 4486. In the case of NGC 1407, the starlight metallicity profile (from spectroscopy) closely matches that of the  red GC metallicity profile (from photometry) in the radial region of overlap (Forbes et al. 2011). For NGC 4486 (M87), there have been several studies of the GC colour gradients and we list their converted metallicity gradients in Table 1. As well as being reasonably consistent with each other, within an individual study the metal-poor and metal-rich GC metallicity 
gradients are the same within the quoted uncertainties. From deep multi-band photometry, Liu et al. (2005) derived a {\it stellar} metallicity gradient between 2.5 and 5 R$_e$ of --0.17 dex per dex (no uncertainty was given). This value is similar to those quoted for NGC 4486's GCs in Table 1, and identical to the --0.17 dex per dex from the GC study of Forte et al. (2012) from 0.2 to 3.5 R$_e$. 

{\it We conclude that, in general,  metal-poor and metal-rich GC subpopulations have the same radial metallicity gradients (see Section 3.1), and that these gradients are similar, if not identical, to those of the galaxy stars over a common radial range. }
It is particularly note-worthy that the metal-poor GCs have a metallicity gradient similar to that of the stars. This suggests that the metal-poor GCs were subject to the same event that established the stellar and metal-rich GC metallicity gradients. This is in stark contrast to other properties for the metal-poor GCs, such as surface density, mean ellipticity, position angle, kinematics etc, that do not appear to have a strong correlation with the equivalent host galaxy stellar property. 


\begin{figure}
      \includegraphics[angle=-90, width=0.5\textwidth]{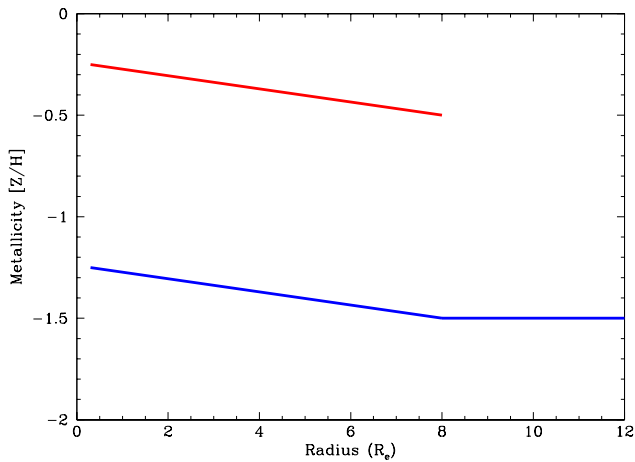}
	    \caption{\label{fig:fake} Schematic representation of 
	    globular cluster metallicity gradients as described in Section  3. The diagram shows the 
	    metallicity gradient of the metal-rich GCs in red. This gradient is similar to that of the stellar metallicity gradient. Approximately 1 dex lower in metallicity, the metallicity gradient for the 
	    metal-poor GCs are shown in blue. At several effective radii (R$_e$)  this radial gradient may give way to a constant metallicity for metal-poor GCs to large radius. 
}

\end{figure}

\begin{figure*}
      \includegraphics[angle=-90, width=0.95\textwidth]{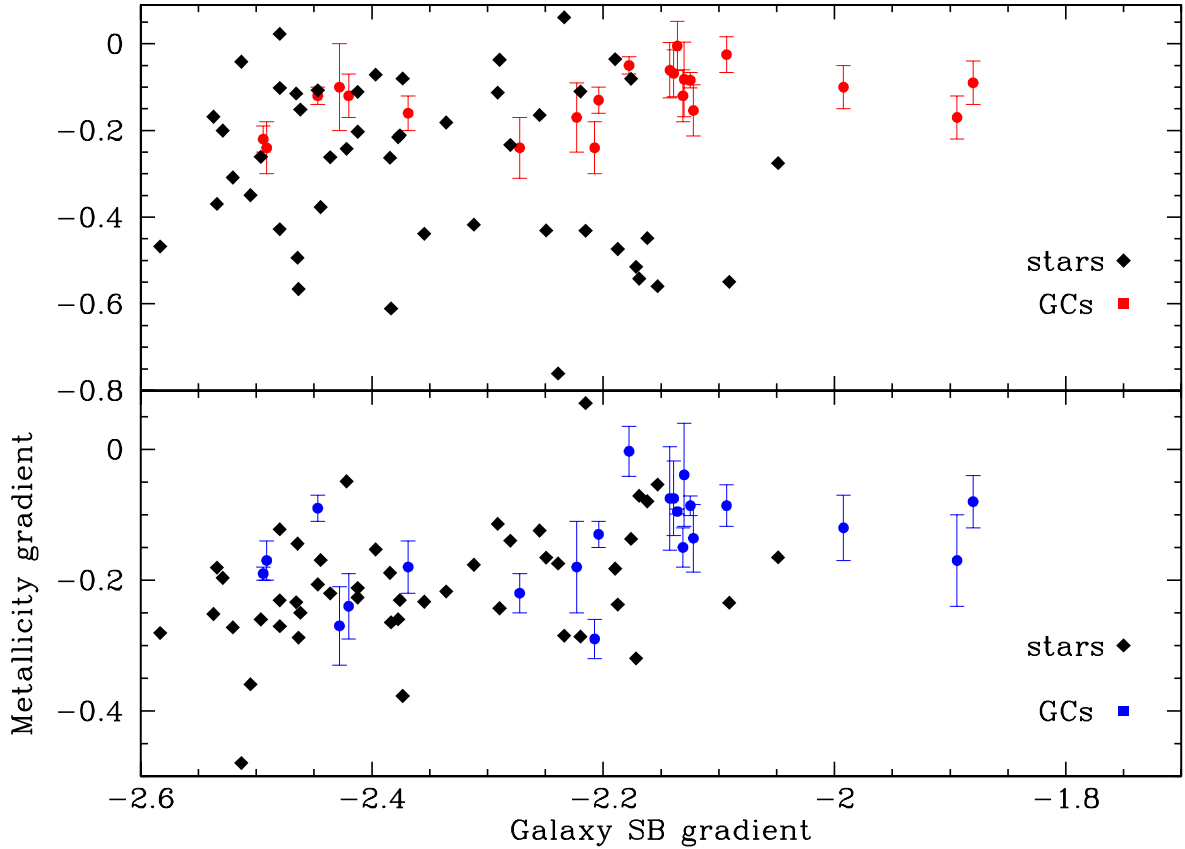}
	    \caption{\label{fig:cook} Globular cluster and stellar metallicity gradients vs galaxy surface 
brightness (SB) gradients. The two panels show the metallicity gradients (in units of dex per dex) for the metal-poor and metal-rich GC 
subpopulations (coloured filled circles) compared to the power-law slope of the host galaxy light profile
measured over the same radial range. The black diamonds show the predicted {\it stellar} metallicity gradients for the accreted (lower panel) and in-situ (upper panel) components vs stellar surface brightness for ETGs from the Magneticum simulations over a fixed radial range of 1--5~R$_e$ (i.e. similar to that of the observed data). 
The observed galaxies and model galaxies occupy a similar parameter space (although the simulations are missing a few galaxies with shallow surface brightness profiles and include galaxies with steeper in-situ stellar gradients than observed for red GCs). 
}

\end{figure*}

\subsection{Outer halo mean metallicity of GC subpopulations}

The `inner halo' GC gradients discussed above, and listed in Table 1, are typically measured out to 5-8 R$_e$.  In some cases, the mean colours of the blue GCs are measured to very large radii. For example in NGC 1407 (Forbes et al. 2011), the red and blue GCs reveal metallicity gradients to about 8 R$_e$. Beyond this radius, in its `outer halo', both GC subpopulations reveal a flat gradient to $\ge$ 15 R$_e$, i.e. a constant colour (although there are few red GCs at such radii). A similar situation of an inner halo colour gradient giving way to a constant mean colour at larger radii is seen for the blue GCs in M87 (Harris 2009b; Strader et al. 2011) and NGC 1399 (Forte et al. 2001).  These outer halo GC colours correspond to mean metallicities of [Z/H] $\sim$ --1.3. Interestingly, we note that Tal \& van Dokkum (2011) found that ETG surface brightness profiles are well represented by a single Sersic fit within 8~R$_e$ but beyond that another component was required. 

The detection of an inner halo gradient and constant colour for the outer halo  requires excellent photometry, a metallicity sensitive filter combination and a large sample of GCs. These criteria largely restrict the discovery to massive ellipticals and their rich GC systems, although we note that a similar feature is seen in the mean (spectrosopic) metallicity of the GCs in the Milky Way with steep gradients within 10 kpc and a constant metallicity beyond that for the metal-poor GCs (Harris 2001). 
 
These observations suggest (at least) 3 distinct categories of GCs: inner red GCs, inner blue GCs and outer blue GCs. The first two reveal similar radial metallicity gradients, whereas the outer halo blue GCs have a constant metallicity indicative of a separate process.
If the outer halo blue GCs have a different origin to the inner halo GCs then we might expect other properties  to vary as a function of galactocentric radius. 
Perhaps the best studied nearby galaxy with a rich GC system is that of NGC 4486 (M87). For this galaxy, Harris (2009b) used photometry from CFHT/Megacam imaging covering a square degree, while Strader et al. (2011) combined spectroscopic and HST-derived size information with a re-analysis of the CFHT/Megacam imaging used by Harris (2009b). 
For reference, at a distance of 16.7 Mpc 1 arcmin corresponds to 4.8 kpc and the effective radius of the galaxy is 7 kpc (Forbes et al. 2017).  
Harris (2009b) found radial gradients in the mean GC red and blue subpopulations out to about 8 R$_e$ ($\sim$12 arcmins). Beyond that, the blue GCs reveal a constant mean colour. The transition from inner gradient to outer constant colour occurs at a similar host galaxy effective radius to that of NGC 1407's GC system (Forbes et al. 2011). 

Strader et al. (2011) published mean position angles (PA) and ellipticities for the red and blue GCs separately in 4 radial bins out to 12 arcmin (8 R$_e$). The mean ellipticity of the blue GCs are the same within the uncertainties at all radii. The mean PA of the blue GCs oscillates around --30 degrees but again with no clear radial trend. So the blue GCs in M87 do not reveal any strong differences in their azimuthal distribution with radius, although the radii probed are all within the 12 arcmin inner halo region for which Harris (2009b) suggests a radial metallicity gradient exists. (We are not aware of any galaxy study that probes the azimuth distribution  of the blue GCs to radii that cover the outer halo region of constant mean blue GC colour.) 
The GC kinematics presented by Strader et al. (2011) do, however, probe beyond 12 arcmins. They find that the blue GCs reveal a kinematic twist, a sharp decline in velocity dispersion and a significant change in kurtosis around 12 arcmin.  This is suggestive of a transition between the inner halo and outer halo blue GCs, and hence a different origin.

Our findings from this Section for the radial metallicity distribution of red and blue GCs in massive early-type galaxies are summarised in a schematic way in Fig.~\ref{fig:fake}. This representative diagram shows the same radial metallicity gradient for the red (metal-rich) GCs and the blue (metal-poor) GCs, only offset by 1 dex in metallicity. We have argued that the stellar metallicity gradient is similar to that of the red GCs. At some large effective radius (8 R$_e$ in the diagram), the blue GC metallicity gradient gives way to an outer halo gradient that is flat, i.e. constant metallicity. We refer the reader to fig. 1 of Forbes et al. (2011) and fig. 10 of Faifer et al. (2011) for real-world examples of our schematic plot, which show not only the mean gradients but also the individual GC data points.

\section{Comparison with Magneticum Simulations}

Although a number of simulations have modelled the formation of GCs in a hierarchical cosmology (e.g. Bekki et al 2008; Muratov \& Gnedin 2010; Li \& Gnedin 2014; Kruijssen 2015; Choksi et al. 2018) with reasonable success in reproducing bimodal metallicity distributions and some global scaling relations, none have predicted the radial colour or metallicity profiles for GC subpopulations (although see Pfeffer et al. 2018 for a first step in modelling the spatial distribution of GC systems in Milky Way-like galaxies). However, simulations of the {\it stars} in massive ETGs are available and, as noted above, GC metallicity gradients tend to be similar to those of their host galaxy {\it stellar} metallicity profiles. Thus it is useful to investigate simulated stellar gradients as a proxy for GC gradients.

The Magneticum hydrodynamical cosmological simulation suite (Dolag et al., in prep.) produces in its highest resolution ($m_\mathrm{gas} = 7.3\times 10^6 M_\odot/h$ and $M_\ast \approx 1.8\times 10^6 M_\odot/h$ as every gas particle can spawn up to four stellar particles) massive ETGs with properties that successfully reproduce observations (e.g. Teklu et al. 2015, for stellar-mass--angular-momentum properties; Remus et al. 2017, for mass-size relations at different redshifts and dark matter properties; Teklu et al. 2017, for satellite fractions and the morphology-density relation; Schulze et al. 2018, for kinematic properties).
The simulation adopts a WMAP7 (Komatsu et al., 2011) $\Lambda$CDM cosmology with $\sigma_8 =0.809$, $h = 0.704$, $\Omega_\Lambda = 0.728$, $\Omega_\mathrm{M} = 0.272$ and $\Omega_\mathrm{B} = 0.0451$, and an initial slope for the power spectrum of $n_\mathrm{s} = 0.963$ (see Hirschmann et al. 2014; Teklu et al. 2015 for more simulation details, Dolag et al. 2017 for details of the metal model and Remus et al. 2017 for the ETG identification criteria).

\begin{figure*}
      \includegraphics[angle=0, width=0.95\textwidth]{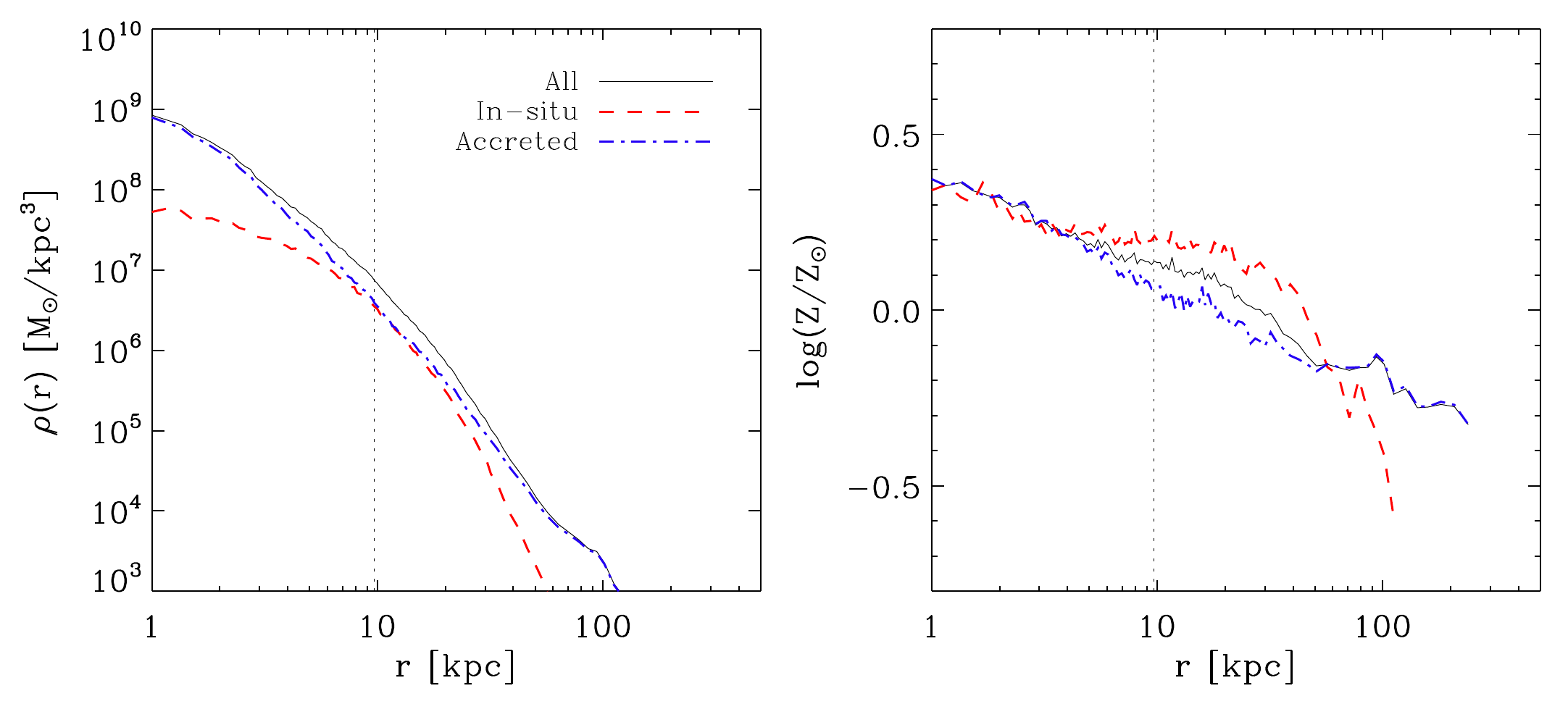}
	    \caption{\label{fig:density} Stellar profiles for the NGC 1407-like model Magneticum ETG. {\it Left:} Radial stellar volume density profiles for the in-situ formed stars (red dashed line) and the accreted stars (blue dash-dotted line), as well as all stars (solid black line). 
	    The vertical line marks the galaxy half-mass radius. For this massive galaxy, accreted stars make the dominant contribution at all radii. 
	    {\it Right:} Radial stellar metallicity profiles for the in-situ, accreted, and all stars, colours as in the left panel.
	    }
\end{figure*}
 
For this work, we select ETGs from the Magneticum simulation in the same mass range as the observed sample presented in Table 1, i.e. $10.9 < \mathrm{log}(M_\ast) < 11.6$ (hereafter Magneticum ETGs). These ETGs are a subset of those presented by Remus et al. (2017),  which should be consulted for further details.   
As the galaxies are triaxial, we calculate the radial edge-on surface density and fit a power-law to the profile in the range of 1--5~R$_{e}$, to emulate observed surface brightness gradients (and we assume a mass-to-light ratio of 1). 
Accordingly, we calculate the stellar metallicity gradients for the accreted and in-situ components separately for each galaxy, over the same fixed radial range of 1--5~R$_e$.




In Fig.  \ref{fig:cook}  we show the GC subpopulation metallicity gradients vs the host galaxy surface brightness gradient calculated over the same radial range as the metallicity gradient. We also show predictions from the Magneticum ETGs for the {\it stellar} metallicity gradient of the in-situ (upper panel) and accreted (lower panel) components vs the stellar surface density gradient.  
It is reassuring that the GC data and the simulations occupy a similar region of parameter space, 
giving us some confidence in using stellar gradients as a proxy for GC gradients. 
Although the gradients measured over the range 1--5~R$_e$ lack some of the flatter surface brightness gradients, we note that shallower gradients can simply be produced if a smaller radial range is measured from the model galaxies, e.g. 1--4~R$_e$. The agreement is particularly good between the metal-poor GCs and the accreted stellar profiles. The simulations produce some relatively steep in-situ stellar metallicity gradients, compared to the range of gradients observed for the metal-rich GCs. This is 
because gas particles that are accreted and then later form stars in the primary galaxy are 
tagged as in-situ stars.  

Next we examine a single ETG from the Magneticum suite in more detail. We picked a 
massive (log (M$_{\ast}$) = 11.6, R$_e$ = 9.7 kpc) model galaxy to resemble NGC 1407 but with no knowledge of its metallicity gradients or assembly history.  NGC 1407 has a 
rich GC system, which has been well-studied beyond 8 R$_e$ as part of the SLUGGS survey (e.g. Forbes et al. 2011).  As such, a model galaxy with global properties similar to NGC 1407 may provide further useful insight. 
In Fig. \ref{fig:density} we show the radial distribution of the stellar volume density of the in-situ and accreted stars for the NGC 1407-like model galaxy at z = 0. This figure shows that, although the model galaxy has a significant in-situ formed component, the stars accreted over cosmic time ultimately dominate at all radii. 
Beyond 60 kpc ($\approx$ 6 R$_e$), accreted stars dominate over in-situ stars by a factor of more than ten in surface density, which leads to a flattening of the overall density profile. The fraction of accreted to total stars at z = 0 in this model galaxy is 70\%. 

Tonini (2013) showed, in her hierarchical cosmology model, that a
log M$_{\ast}$ = 11 galaxy with a 70\% accretion fraction of the total mass (i.e. similar to our NGC1407-like model) should have a bimodal GC metallicity distribution. However, if the accretion fraction is lowered in her model to 50\%, then the GC metallicity distribution no longer resembles those observed, being depleted in metal-poor GCs.
We note that a galaxy of log M$_{\ast}$ = 11 today contains roughly equal fractions of metal-rich and metal-poor GCs (Peng et al. 2006).

Fig. \ref{fig:density} also shows the radial metallicity profiles of the in-situ and accreted stars in the NGC 1407-like model. The in-situ formed stars have a relatively flat inner metallicity gradient but it drops steeply beyond 60 kpc. Beyond 100~kpc, low number statistics make the profile unreliable. Interestingly, within the innermost region  (approximately within the half-mass radius), the metallicities of the in-situ and accreted components are identical, while at larger radii the in-situ component shows the expected behaviour of being more metal-rich than the accreted component. 

\begin{figure}
      \includegraphics[angle=0, width=0.5\textwidth]{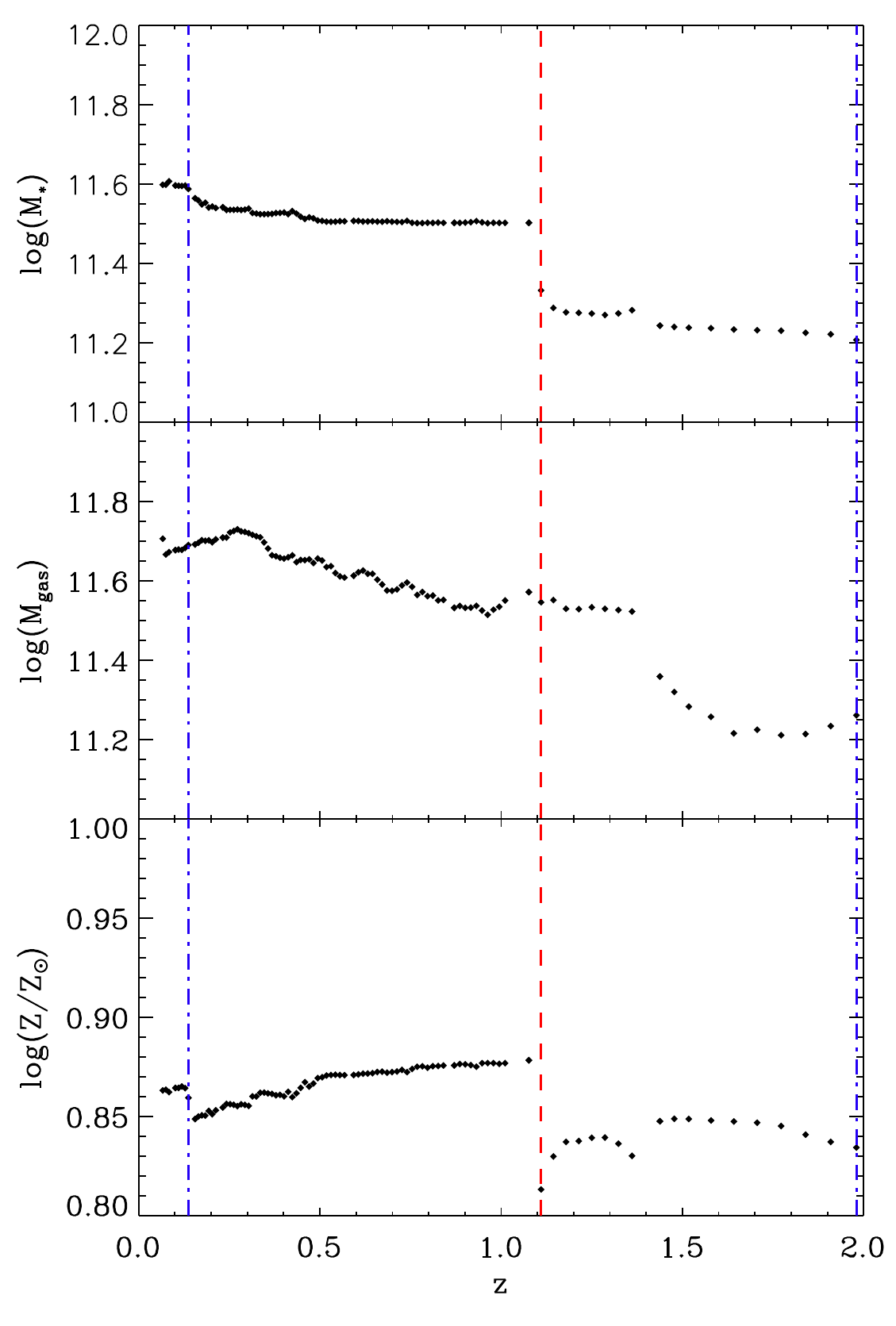}
	    \caption{\label{fig:history} Mass assembly history of stars in the Magneticum NGC 1407-like model. {\it Top:} Stellar mass evolution; {\it Middle:} Total gas mass evolution (which is dominated by hot gas); {\it Lower:} Total stellar metallicity. Vertical lines represent merger events, i.e. a 88:1 merger at z = 2, a 2:1 merger at z = 1.1, and  a 57:1 merger at z = 0.14. 	    	    
	    }
\end{figure}
To understand the behaviour of the in-situ and accreted components in both density and metallicity, it is important to understand the formation history of this galaxy:
Fig. \ref{fig:history} shows the mass assembly history for the NGC 1407-like model galaxy, for its stellar component (upper panel), its gas component (middle panel), and its mean stellar metallicity (lower panel). Before z = 2 (i.e. more than 10 Gyr ago), the galaxy formed most of its stars in a disk-like configuration. At redshift $z \approx 2$, it undergoes a mini-merger with a mass ratio of 88:1. Such an event would not be expected to significantly change the GC system of the main galaxy, and we do not see any impact on its kinematics. 

At $z \approx 1.4$, the galaxy experiences a gas-rich merger, with a stellar mass ratio below 100:1 but a gas ratio that is nearly equal to the gas already present in the host. The gas accreted in this event is metal-poor, effectively lowering the mean metallicity.

At z = 1.1 (8.2 Gyr ago), the galaxy merges with another galaxy of similar mass in a 2:1 major merger. This event significantly raises the metallicity and the stellar mass of the remnant. 
Major mergers are thought to be the main pathway for galaxies with $\mathrm{log}(M_{\ast}) > 11.3$ (Bernardi et al. 2011) to deliver stars (and any gas) into the heart of the host galaxy (Rodriguez-Gomez et al. 2016; Amorisco 2017).  
The massive satellite (with $\mathrm{log}(M_{\ast}) \sim 11$) would have formed its own system of metal-rich and metal-poor GCs at earlier times (today such a galaxy would have equal subpopulations of GCs; Peng et al. 2006). Its metal-rich GCs would be of similar mean metallicity as the galaxy mass-metallicity relation is relatively flat for massive galaxies (Erb et al. 2006; Shapiro et al. 2010). Similarly, it should possess a significant number of metal-poor GCs as well. 
Thus, both galaxies would have hosted metal-rich and metal-poor GC subpopulations before their encounter.  

In recent times (z = 0.14) the galaxy suffers a minor merger of ratio 57:1, which is gas-poor and metal rich. This galaxy of mass $\mathrm{log}(M_{\ast}) < 10$ would be dominated ($>$ 2/3) by metal-poor GCs (Peng et al. 2006). Such mergers of low mass satellites mostly contribute mass to the outer parts of galaxies (e.g. Karademir et al., 2018), and may be responsible for flattening both the surface brightness and stellar metallicity profiles in the outer halo, and for generating the near constant (blue)  colour of the outer halo GCs. 

The final galaxy, at z = 0, has a total metallicity $\sim$10\% of the solar metallicity and a stellar mass of $\mathrm{log}(M_{\ast}) = 11.6$. 
It also has a similar mass in gas, most of which is contained in a diffuse hot gas corona. 

\begin{figure}
      \includegraphics[angle=0, width=0.5\textwidth]{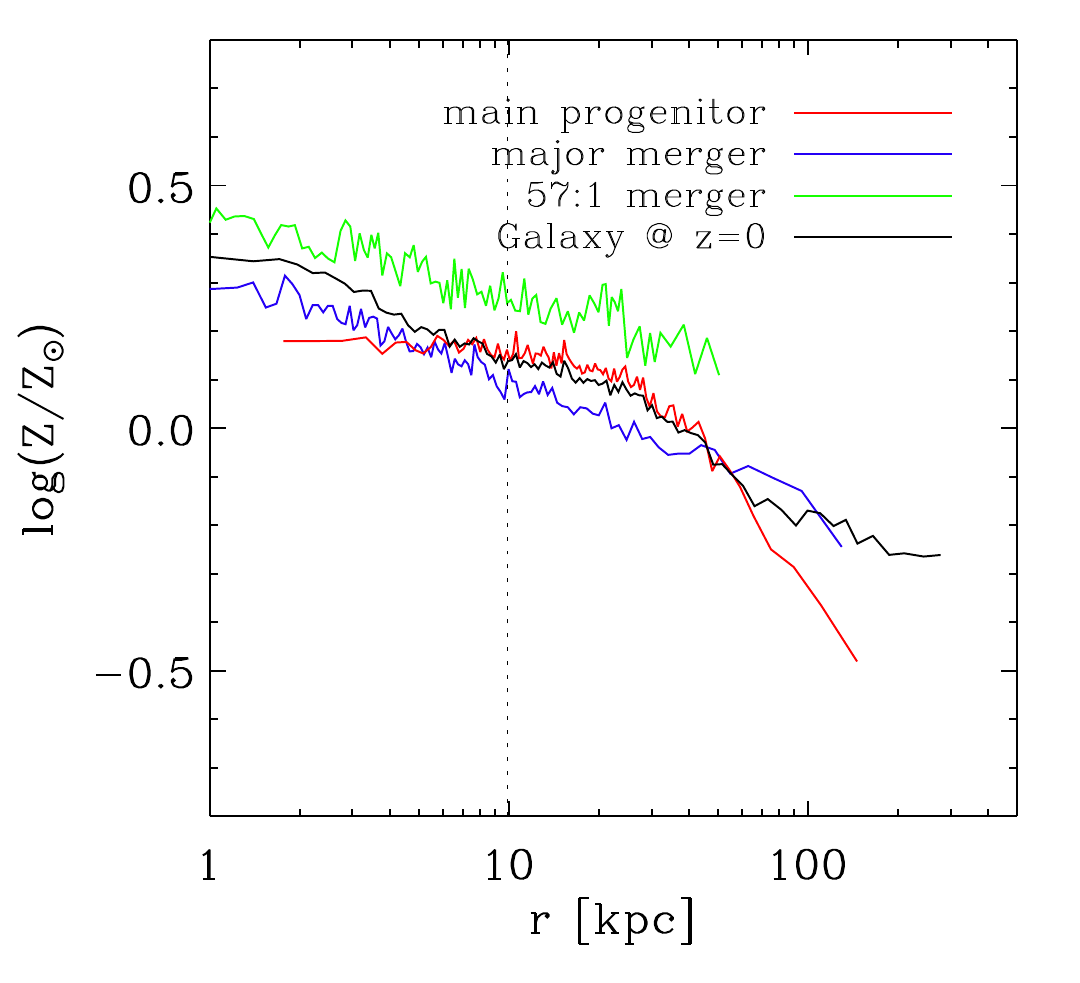}
	    \caption{\label{fig:profiles} Stellar metallicity profiles for the  Magneticum NGC 1407-like model.  The red line shows the stars formed in-situ within the main host galaxy, the blue line is the accreted stars from the 3:1 major merger and the green line from the minor merger. The black line shows the metallicity profile for all stars at z = 0 in the final galaxy. The vertical line shows the half-mass radius of the final galaxy. 
	    	    }
\end{figure}

To understand the contribution of the different merger events to the radial metallicity profile, we show in Fig. \ref{fig:profiles} the radial metallicity profiles of the stars that were already in the host prior to the major merger (red line), the stars that came in through the major merger (blue line), and those stars that came in through the 57:1 merger at $z \approx 0.1$ (green line). 
Note that there are stars formed in-situ from gas inside the galaxy after the major merger at $z\approx 1.1$, which are only included in the total stellar metallicity profile and not in any of the three separate components, and that these new stars cause the difference between the final galaxy (black) and the major merger (blue) curves in the inner regions  of the galaxy.

The in-situ component and massive satellite have almost identical metallicity gradients between 1 and 5 R$_e$ of --0.31 and --0.33 dex per dex respectively. The plot also shows that although the final galaxy at z = 0 has a similar metallicity gradient to the  in-situ component, over this radial range, the outer gradient is significantly flatter than the original in-situ gradient. 

This single massive early-type model galaxy was chosen to have similar global properties to NGC 1407, independent of its mass assembly. Although the final mass is a key determinant 
of assembly history, the process is a stochastic one and a number of pathways are possible for mass growth. This will lead to a range of metallicity profiles. To see how this influences the resulting slope properties, we now examine the full set of Magneticum ETGs that match the stellar mass range of our observed galaxy sample. 

\begin{figure*}
      \includegraphics[angle=0, width=0.95\textwidth]{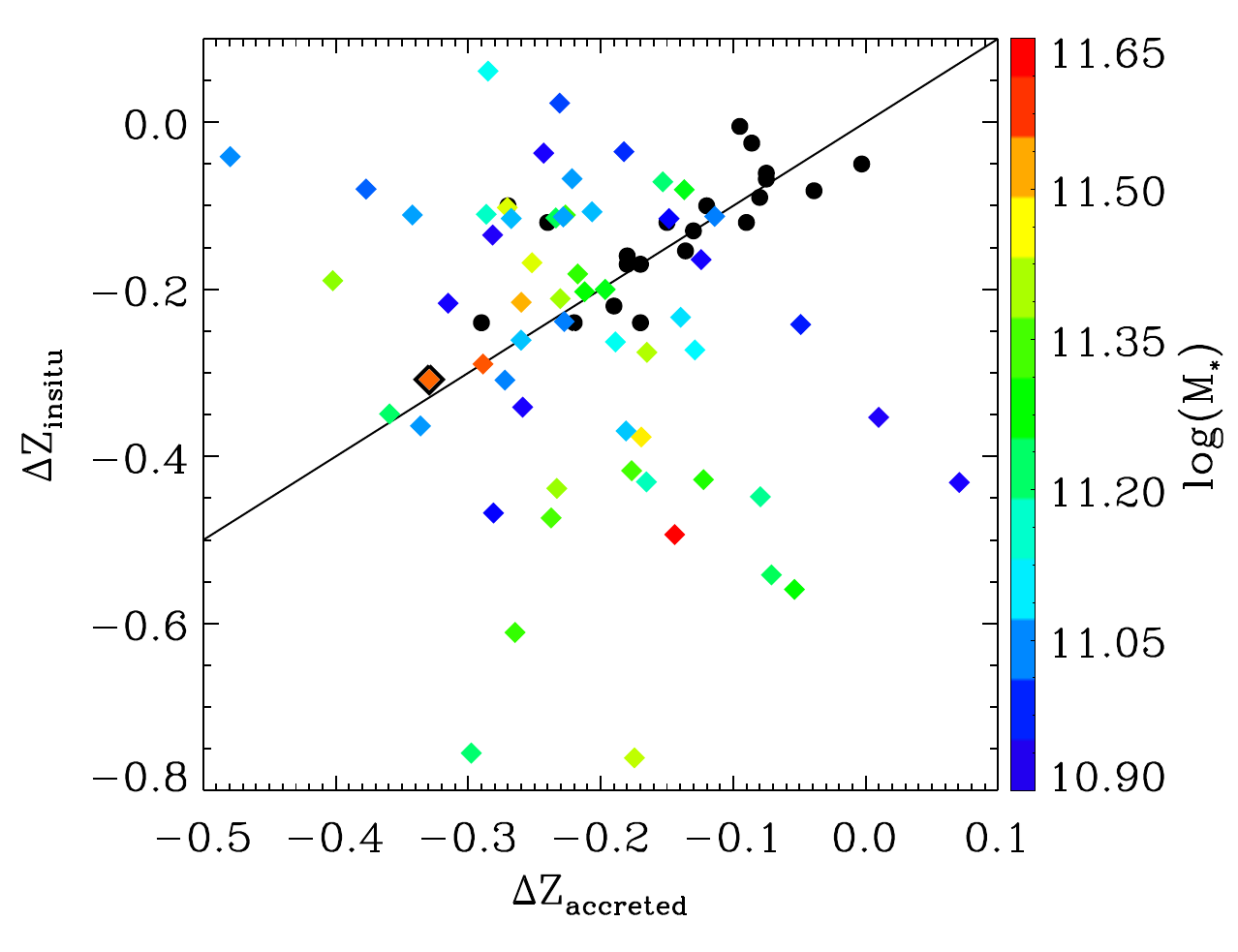}
	    \caption{\label{fig:mass} Stellar accreted vs in-situ 
	    metallicity gradients (1--5~R$_e$) for Magenticum ETGs. Coloured symbols are coded by stellar mass, which matches the mass range of observed galaxies in Table 1. The large orange data point shows the NGC 1407-like model galaxy highlighted in Figures 3, 4 and 5. Observed GC system metallicity gradients are shown by black circles, with the gradient for the blue GCs shown on the x axis and the red GCs on the y axis. The solid line shows the one-to-one relation. 
	    	    }

\end{figure*}


Fig. \ref{fig:mass}  shows the stellar metallicity gradients of the in-situ and accreted components for the Magneticum ETGs over a mass range similar to our sample galaxies in Table 1. The simulated galaxies are colour coded by their stellar mass.
The gradient radial range is fixed to be 1 to 5 R$_e$ (to roughly match the typical range of our GC metallicity gradients). The plot shows that the in-situ and accreted stellar metallicity gradients are similar for many galaxies (as seen in Fig. \ref{fig:density} for the NGC 1407-like model). As shown in Fig.~\ref{fig:cook}, the simulations have a wider range in their stellar metallicity gradients than the GCs. This is partially due to the steeper in-situ gradients produced in the simulation. If the steep in-situ gradients (e.g. steeper than --0.4 dex per dex) were excluded, then the 
simulated gradients would scatter fairly evenly about the unity line. 
We note that there is a weak trend for these steeper in-situ gradients to be found in the more massive model galaxies, which indicates a stronger influence from newly formed stars in galaxy centres than seen in the red GC subpopulation gradients.

\section{The origin of GC subpopulations in massive ETGs: an overview}

We identify metal-poor and metal-rich `inner halo' GCs and `outer halo' metal-poor GCs based on their radial metallicity gradients, with a transition occurring at several galaxy effective radii (tens of kpc). 
We now attempt to include our findings in the wider context of two-phase galaxy formation (e.g. Oser et al. 2010) and propose the following scenario:

During the first phase of massive galaxy formation, which resembles a gaseous monolithic-like collapse (Cook et al. 2016), in-situ stars and metal-rich GCs form in gaseous disks and in gaseous mergers at early times (Muratov \& Gnedin 2010; Li \& Gnedin 2014; Kruijssen 2015; Choski et al. 2018). 
These  stars and metal-rich GCs form with a similar mean metallicity as they are both set at birth by the depth of the galaxy potential well, i.e. the mass-metallicity relation at that redshift (Shapiro et al. 2010). 
However at these relatively high metallicities, the combination of formation efficiency and destruction result in few metal-rich GCs compared to the formation of field stars, i.e. a very low GC-to-star ratio (Lamers et al. 2017). 

Prior to the formation of these metal-rich GCs, metal-poor GCs will have formed in low mass galaxies with 
a relatively high efficiency of formation, i.e. the low metallicity gas results in a very high GC-to-star ratio (Lamers et al. 2017). On average, metal-poor GCs will have older ages than metal-rich ones. 
Thus two subpopulations of GCs, with different mean metallicities and mean ages, are established. The metallicity differences are largely a function of the mass of the host galaxy in which they formed. 

The second phase of galaxy growth occurs by accretion.
A merger involving a similar, or  lower, mass galaxy will bring in, on average, more metal-poor than metal-rich GCs (Peng et al. 2006). 
Mergers redistribute the stars (and the galaxy potential via violent relaxation), and may flatten any existing (in-situ) metallicity gradient (Di Matteo et al. 2009). 
So originally the metal-poor GCs formed first in low mass galaxies and were then later accreted by a larger galaxy. The metal-rich GCs formed in-situ within the primary `seed' galaxy (and in massive satellites). 

In the picture outlined above, we suggest that the radial metallicity profiles of both the 
metal-rich and metal-poor GCs, in the inner halos of today's massive galaxies, are the result of relatively major mergers during the second phase of galaxy growth (i.e. z $\le$ 2). 
However, there is {\it not} a one-to-one correspondence between metal-rich GCs and the in-situ phase of galaxy formation. Similarly, metal-poor GCs are not all accreted, although their in-situ fraction is very low. 
Major mergers, in the accretion phase, will bring in {\it both} metal-rich and metal-poor GCs. 

Major mergers of massive satellites will tend to deposit GCs deep into the host galaxy inner regions due to their higher binding energies, whereas low mass satellites will be tidally disrupted at large radii and hence deposit their stars, and predominately blue GCs, into the outer halo region (Karademir et al. 2018). 
As low mass galaxies tend to have GCs of a similar mean colour, the deposited GCs will result in a near constant mean colour (metallicity) in the outer halo regions as described in Section 3.3. 
The binding energy argument may also explain the observation that the surface density distribution of blue GCs tends to be more radially extended than the red GCs. 



A number of predictions from this scenario can be directly tested:\\
$\bullet$ Galaxies that have only undergone the in-situ phase of galaxy formation with few/no subsequent mergers (e.g. relic galaxies; Ferre-Mateu et al. 2015, 2017) should be dominated by red GCs. On the other hand, galaxies that have had strong levels of accretion should be dominated by blue GCs. Examples of such systems may exist, e.g. NGC 1277 which is largely devoid of blue GCs (Beasley et al. 2018) and NGC 4874 which reveals only a unimodal distribution of blue GCs (Harris et al. 2000). \\
$\bullet$ The outer halo blue GCs may have a different mean age, [$\alpha$/Fe] ratio, orbital and spatial properties to inner halo blue GCs.\\
$\bullet$ The transition from an inner halo colour gradient to an outer halo constant colour is the transition from major to minor mergers dominating the accretion. A change in the galaxy surface brightness and galaxy colour profiles are also expected at this transition radius.\\
$\bullet$ Higher mass galaxies (that have experienced more merger/accretion activity) may have shallower GC and stellar metallicity gradients, more extended GC systems, higher blue-to-red GC ratios and higher GC specific frequencies.\\
$\bullet$ As the blue GCs are accreted in major and minor mergers it may be possible to detect a change in the radial surface density slope at the transition from GCs contributed by mostly major to those from minor mergers. For low mass galaxies, that have a lower fraction of accreted material, the slopes of the blue and red surface density profiles may be more similar.

\section{Conclusions and Future Work}

The blue (metal-poor) and red (metal-rich) GC subpopulations around massive early-type galaxies reveal radial colour (metallicity) gradients within several effective radii. We have collated such colour gradients from the literature and homogeneously converted them into metallicity gradients. We find that in a given galaxy, the metallicity gradients of the blue and red GC subpopulations are the same within the measurement uncertainty. 
Furthermore, these GC metallicity gradients are similar to those of the galaxy stellar metallicity gradient when compared over the same range in galactocentric radii. 

As current simulations do not provide predictions for the radial metallicity distribution of GC systems, we investigate hydrodynamical models that track the {\it stellar} component as a proxy for the GCs.  We find that the GC subpopulation gradients as a function of their host galaxy surface brightness profile gradient have similar values to those of model galaxies from the Magneticum simulation. 

Again using Magneticum, we follow the assembly history of the {\it stars} in a single massive early-type galaxy. This model galaxy undergoes mergers of varying mass ratios during its `accretion phase'  of z $\le$ 2. The original in-situ metallicity gradient and metallicity gradient for the massive satellite that is accreted in a major merger,  are 
almost identical between 1 and 5 effective radii (i.e. a similar range to that measured for GC gradients). Furthermore, the gradient of the final galaxy at z = 0 over this radial range is also similar. 

We speculate that the metal-poor and metal-rich  GC subpopulation gradients are the same in a given galaxy because they are the result of the same mergers that give rise to near identical in-situ, accreted and final stellar gradients. The outer halo regions, which reveal flat GC metallicity gradients, are likely the result of recent minor mergers with low mass satellites. 


Recently, star cluster formation has been incorporated into the EAGLE simulation for a dozen Milky Way like galaxies in order to examine their GC systems (Pfeffer et al. 2018). Although the simulation currently over produces the number of metal-rich GCs at intermediate galactocentric radii, it is an excellent first attempt at predicting the spatial properties of GC systems in a cosmological context. When the simulations are extended to giant ellipticals, a key observational property to reproduce will be the similarity of the metal-poor and metal-rich GC subpopulation metallicity gradients found in this work. Such simulations should also be in an ideal position to probe the full assembly history of individual metal-rich and metal-poor GCs.

\section{Acknowledgements}

DAF thanks the ARC for financial support via DP130100388. The  Magneticum Pathfinder simulations were partially performed at the
Leibniz-Rechenzentrum with CPU time assigned to the Project pr86re. This work was supported by the DFG Cluster of Excellence Origin and Structure of the Universe. We are especially grateful for the support by M. Petkova through the Computational Center for Particle and Astrophysics (C2PAP). Both DAF and RSR thank DAAD for financial support. 
We thank A. Ferre-Mateu and K. Dolag for helpful discussions. Finally, we thank the referee for several useful suggestions.

\section{References}

Alabi A.~B., et al., 2016, MNRAS, 460, 3838\\
Amorisco N.~C., 2017, MNRAS, 464, 2882\\
Arnold J.~A., Romanowsky A.~J., Brodie J.~P., Chomiuk L., Spitler L.~R., Strader J., Benson A.~J., Forbes D.~A., 2011, ApJ, 736, L26 \\
Bassino L.~P., Faifer F.~R., Forte J.~C., Dirsch B., Richtler T., Geisler D., Schuberth Y., 2006, A\&A, 451, 789 
Beasley M.~A., Trujillo I., Leaman R., Montes M., 2018, Natur, 555, 483 \\
Bekki K., Yahagi H., Nagashima M., Forbes D.~A., 2008, MNRAS, 387, 1131 \\
Bernardi M., Roche N., Shankar F., Sheth R.~K., 2011, MNRAS, 412, L6 \\
Blom C., Spitler L.~R., Forbes D.~A., 2012, MNRAS, 420, 37 \\
Boylan-Kolchin M., 2017, MNRAS, 472, 3120 \\
Brodie J.~P., Strader J., 2006, ARA\&A, 44, 193\\
Brodie J.~P., Usher C., Conroy C., Strader J., Arnold J.~A., Forbes D.~A., Romanowsky A.~J., 2012, ApJ, 759, L33 \\
Brodie J.~P., et al., 2014, ApJ, 796, 52 \\
C{\^o}t{\'e} P., Marzke R.~O., West M.~J., 1998, ApJ, 501, 554 \\
Choksi N., Gnedin O., Li H., 2018, arXiv, arXiv:1801.03515 \\
Coccato L., Gerhard O., Arnaboldi M., 2010, MNRAS, 407, L26 \\
Cook B.~A., Conroy C., Pillepich A., Rodriguez-Gomez V., Hernquist L., 2016, ApJ, 833, 158 \\
Di Matteo P., Pipino A., Lehnert M.~D., Combes F., Semelin B., 2009, A\&A, 499, 427 \\
Dolag, K., Mevius, E., Remus, R.-S.\ 2017, Galaxies, 5, 35\\
Erb D.~K., Shapley A.~E., Pettini M., Steidel C.~C., Reddy N.~A., Adelberger K.~L., 2006, ApJ, 644, 813 \\
Faifer F.~R., et al., 2011, MNRAS, 416, 155 \\
Forbes D.~A., Brodie J.~P., Grillmair C.~J., 1997, AJ, 113, 1652 \\
Forbes D.~A., Forte J.~C., 2001, MNRAS, 322, 257 \\
Forbes D.~A., Bridges T., 2010, MNRAS, 404, 1203 \\
Forbes D.~A., Spitler L.~R., Strader J., Romanowsky A.~J., Brodie J.~P., Foster C., 2011, MNRAS, 413, 2943 \\
Forbes D.~A., Ponman T., O'Sullivan E., 2012, MNRAS, 425, 66 \\
Forbes D.~A., Alabi A., Romanowsky A.~J., Brodie J.~P., Strader J., Usher C., Pota V., 2016, 
MNRAS, 458, L44 \\
Forbes D.~A., 2017, MNRAS, 472, L104 \\
Forbes D.~A., Sinpetru L., Savorgnan G., Romanowsky A.~J., Usher C., Brodie J., 2017, MNRAS, 464, 4611 \\
Forte J.~C., Geisler D., Ostrov P.~G., Piatti A.~E., Gieren W., 2001, AJ, 121, 1992 \\
Forte J.~C., Vega E.~I., Faifer F., 2012, MNRAS, 421, 635 \\
Ferr{\'e}-Mateu A., Mezcua M., Trujillo I., Balcells M., van den Bosch R.~C.~E., 2015, ApJ, 808, 79 \\
Ferr{\'e}-Mateu A., Trujillo I., Mart{\'{\i}}n-Navarro I., Vazdekis A., Mezcua M., Balcells M., Dom{\'{\i}}nguez L., 2017, MNRAS, 467, 1929 \\
Graham, A. W., 2013, Planets, Stars and Stellar Systems Vol. 6, by Oswalt, Terry D.; Keel, William C., ISBN 978-94-007-5608-3. Springer Science+Business Media Dordrecht, 2013, p. 91\\
Geisler D., Lee M.~G., Kim E., 1996, AJ, 111, 1529 \\
Goudfrooij P., Kruijssen J.~M.~D., 2013, ApJ, 762, 107 \\
Hargis J.~R., Rhode K.~L., 2014, ApJ, 796, 62 \\
Harris W.~E., Kavelaars J.~J., Hanes D.~A., Hesser J.~E., Pritchet C.~J., 2000, ApJ, 533, 137 \\
Harris, W., 2001, Star Clusters Saas Fee Advanced Courses, Springer-Verlag, ed. L. Labhardt and B. Binggeli.\\
Harris W.~E., 2009a, ApJ, 699, 254\\
Harris W.~E., 2009b, ApJ, 703, 939 \\
Harris W.~E., Kavelaars J.~J., Hanes D.~A., Pritchet C.~J., Baum W.~A., 2009, AJ, 137, 3314 \\
Harris W.~E., Harris G.~L., Hudson M.~J., 2015, ApJ, 806, 36 \\
Hirschmann M., Dolag K., Saro A., Bachmann L., Borgani S., \& Burkert A.,
2014, MNRAS 442, 2304\\
Hirschmann M., Naab T., Ostriker J.~P., Forbes D.~A., Duc P.-A., Dav{\'e} R., Oser L., Karabal E., 2015, MNRAS, 449, 528 \\
Hudson M.~J., Harris G.~L., Harris W.~E., 2014, ApJ, 787, L5 \\
Hudson M.~J., Robison, B., 2018, MNRAS, in press\\
Karademir G.S., Remus R.-S., Steinwandel U., Dolag K., Hoffmann T.L.,
Moster B.P., \& Burkert a., 2018, submitted to MNRAS\\
Kartha S.~S., Forbes D.~A., Spitler L.~R., Romanowsky A.~J., Arnold J.~A., 
Brodie J.~P., 2014, MNRAS, 437, 273 \\
Kartha S.~S., Forbes D.~A., Spitler L.~R., Romanowsky A.~J., Arnold J.~A., Brodie J.~P., 2016, MNRAS, 457, 1702 \\
Komatsu E., et al., 2011, ApJS, 192, 18 \\
Kruijssen J.~M.~D., 2015, MNRAS, 454, 1658 \\
Lamers H.~J.~G.~L.~M., Kruijssen J.~M.~D., Bastian N., Rejkuba M., Hilker M., Kissler-Patig M., 2017, A\&A, 606, A85 \\
Law D.~R., Majewski S.~R., 2010, ApJ, 718, 1128 \\
Lee, M., Kim, E., Giesler, D., 1998, AJ, 115, 947\\ 
Lee M.~G., Park H.~S., Kim E., Hwang H.~S., Kim S.~C., Geisler D., 2008, ApJ, 682, 135\\
Li H., Gnedin O.~Y., 2014, ApJ, 796, 10 \\
Liu Y., Zhou X., Ma J., Wu H., Yang Y., Li J., Chen J., 2005, AJ, 129, 2628 \\
Liu C., Peng E.~W., Jord{\'a}n A., Ferrarese L., Blakeslee J.~P., C{\^o}t{\'e} P., Mei S., 2011, ApJ, 728, 116 \\
Mackey A.~D., Gilmore G.~F., 2003, MNRAS, 345, 747 \\
Muratov A.~L., Gnedin O.~Y., 2010, ApJ, 718, 1266 \\
Naab T., et al., 2014, MNRAS, 444, 3357 \\
Oser L., Ostriker J.~P., Naab T., Johansson P.~H., Burkert A., 2010, ApJ, 725, 2312 \\
Park H.~S., Lee M.~G., 2013, ApJ, 773, L27 \\
Pastorello N., et al., 2015, MNRAS, 451, 2625 \\
Peng E.~W., et al., 2006, ApJ, 639, 95 \\
Pfeffer J., Kruijssen J.~M.~D., Crain R.~A., Bastian N., 2018, MNRAS, 475, 4309 \\
Pillepich A., et al., 2014, MNRAS, 444, 237 \\
Pota V., et al., 2013, MNRAS, 428, 389 \\
Remus R.-S., Burkert A., Dolag K., Johansson P.~H., Naab T., Oser L., Thomas J., 2013, ApJ, 766, 71 \\
Remus R.-S., Dolag K., Naab T., Burkert A., Hirschmann M., Hoffmann T.~L., Johansson P.~H., 2017, MNRAS, 464, 3742 \\
Rodriguez-Gomez V., et al., 2016, MNRAS, 458, 2371 \\
Shapiro K.~L., Genzel R., F{\"o}rster Schreiber N.~M., 2010, MNRAS, 403, L36 \\
Spitler L.~R., Forbes D.~A., 2009, MNRAS, 392, L1 \\
Schulze, F., Remus R.-S., Dolag K., Burkert A., Emsellem, E., van de Ven, G., 2018, arXiv:1802.01583\\
Strader J., et al., 2011, ApJS, 197, 33 \\
Tal T., van Dokkum P.~G., 2011, ApJ, 731, 89 \\
Teklu, A.~F., Remus, R.-S., Dolag K., Beck M.A., Burkert, A., Schmidt A.S., Schulze, F., Steinborn L.K., 2015, ApJ, 812, 29\\
Teklu, A.~F., Remus, R.-S., Dolag, K., Burkert, A., 2017, MNRAS 472 4769\\
Tonini C., 2013, ApJ, 762, 39 \\
Usher C., et al., 2012, MNRAS, 426, 1475 \\
Usher C., Forbes D.~A., Spitler L.~R., Brodie J.~P., Romanowsky A.~J., 
Strader J., Woodley K.~A., 2013, MNRAS, 436, 1172 \\
Usher, C., 2014, PhD Thesis, Swinburne University

\end{document}